\documentclass[aps,prd,a4paper,superscriptaddress,nofootinbib,
showpacs,twocolumn,showkeys,amsfonts,amssymb,amsmath]{revtex4-1}

\usepackage{amssymb,latexsym}
\usepackage{amsmath, amsthm}
\usepackage{amscd}
\usepackage{times}
\usepackage{epsfig}
\usepackage{psfrag}
\usepackage{graphicx}

\begin{document}
\title{Scalar field collapse and cosmic censorship}

\author{Rituparno Goswami} \email{Rituparno.Goswami@uct.ac.za}
\affiliation{ACGC and Department of Mathematics and Applied
  Mathematics,
University of Cape Town, Cape Town, South Africa}
\author{Pankaj S. Joshi} \email{psj@tifr.res.in}
\affiliation{Tata Institute of Fundamental Research, Homi Bhabha Road,
Mumbai 400005, India}
\author{Daniele Malafarina} \email{daniele.malafarina@polimi.it}
\affiliation{Tata Institute of Fundamental Research, Homi Bhabha Road,
Mumbai 400005, India}
\swapnumbers

\begin{abstract}
We analyze here the final fate of complete gravitational collapse
of a massless scalar field within general relativity. A class of
dynamical solutions with initial data close to the
Friedmann-Lema{\^i}tre-Robertson-Walker (FLRW) collapse model
is explicitly given and the Einstein equations are integrated
numerically in a neighborhood of the center.
We show that the initial data space is evenly divided
between the dynamical evolutions that terminate in a black hole
final state and those that produce a locally naked singularity.
We comment on the genericity aspects of the collapse end-states
and the connection to cosmic censorship conjecture is
pointed out.
\end{abstract}
\maketitle

The Cosmic Censorship Conjecture (CCC)
\cite{CCC}
has been one of the most prominent open problems
of modern general relativity. The original idea of
Penrose was that the evolution of a generic and physically
realistic matter cloud, subject to Einstein equations,
would never lead to spacetime singularities that are visible
to faraway observers. Over the last few decades much
work has been done on CCC, with multiple formulations being
proposed towards a possible proof, and on what is meant
by genericity of a certain collapse outcome,
a concept for which there is no precise definition
in gravitation theory today.

While a proof of the CCC has not been found
despite many efforts, and its very formulation has been
questioned, many models of physically
realistic gravitational collapse leading to the
formation of naked singularities have been obtained
and studied
(see for example 
\cite{NS}
and references therein).
These models mainly deal with fluids
such as dust, perfect fluids, and other more general
matter fields, which have been used widely in astrophysical
considerations. These are typically described by
averaged properties where the microscopic structure of
the fluid itself may have been neglected in the very
final stages of collapse
(see, however, \cite{Hagadorn}).
In any case, it is not clear whether the fluid
approximation will continue to hold till the ultra-high
densities that characterize the singularity are reached,
or what is the final form of matter and equation of
state valid at such ultra-high densities.

On the other hand, much work has also been done
on scalar field collapse
(\cite{scalar}-\cite{scalar4}).
Scalar fields are fundamental fields satisfying
the Klein-Gordon equation and therefore, despite the
fact that they have never been observed in nature,
constitute a perfect candidate for a fundamental
field whose validity will hold at any scale.
Also, as shown by Choptuik
\cite{scalar2},
the black hole threshold in the space of
initial data for classes of massless scalar fields
show universality and power-law
scaling of the black hole mass, which correspond to
critical phenomena. These are explained by the existence
of exact solutions which are attractors within the black hole
threshold and are typically self-similar. The critical
phenomena evolve a smooth initial data to arbitrarily
large curvatures visible from infinity, via the formation
of infinite time naked singularities that form at the
boundary between the continual collapse to a black
hole and the region where collapse reverses, turning to
a dispersal (see
\cite{scalar4}
for details). The genericity and stability
aspects of naked singularities arising in certain
classes of collapsing massless scalar fields were
investigated by Christodoulou
\cite{scalar1},
showing these to be non-generic within
the context of a certain parent space.

We investigate here complete collapse of a class of
spherically symmetric massless scalar fields within
general relativity. We show explicitly that there exist sets
of non-zero measure of initial data, which are arbitrarily
close to the FLRW
homogeneous collapse scenario when the collapse begins, and
which terminate in a naked singularity. We note that the
naked singularities considered here are of a different nature
compared to the infinite time singularity in the Choptuik
like analysis in the fact that in our models the spacetime
is no longer assumed to be self-similar, and these are
much similar to the central singularities of fluid models
that may be locally visible.

The energy-momentum tensor for a massless scalar field
$\phi(r,t)$ is given by,
\begin{equation}
    T_{ab}=\phi_{;a}\phi_{;b}-\frac{1}{2}g_{ab}(\phi_{;c}\phi_{;d}g^{cd}) \; .
\end{equation}
We consider here the class of scalar fields which admit
one timelike and three spacelike eigenvectors
(see e.g.
\cite{HE}),
and consider the continual
gravitational collapse of such a field.
The FLRW scalar field collapse is an example of
such a scenario. Then $T_{ab}$ can be written in
comoving coordinates and becomes diagonal. The spherically
symmetric line element then depends on three functions
of $t$ and $r$ (which are the
comoving time and radius respectively), which are
$g_{00}=-e^{2\nu(r,t)}$, $g_{11}=e^{2\psi(r,t)}$ and
$g_{22}=g_{33}\sin^2\theta=R(r,t)^2$.
As we can always choose here a frame along the eigenvectors
of the energy-momentum tensor of the scalar field, we have
$T_0^1=0$. We then get $\dot{\phi}\phi'=0$. We shall consider
here the class of scalar fields for which $\phi(t)\neq 0$
~\cite{scalar}.
Then the components of the energy momentum tensor become
$T^a_b=\frac{1}{2}e^{-2\nu}\dot{\phi}^2[-1,1,1,1]$,
from which we see that the matter behaves like a stiff
perfect fluid with an equation of state $p=\rho$, with
$\rho$ given by the equation above.

The system of Einstein equations in units where $8\pi G=c=1$,
can be written as
\begin{eqnarray}\label{rho}
  F' &=& \rho R^2R' \; , \\ \label{p}
  \dot{F} &=& -pR^2\dot{R} \; , \\ \label{nu}
  0&=& \partial_t(R^2e^{\psi-\nu}\dot{\phi}) \; , \\ \label{G}
  \dot{G}R' &=& 2\nu'{\dot{R}}G \; .
\end{eqnarray}
where primed and dotted quantities denote partial
derivatives with respect to $r$ and $t$ respectively, and
we have defined $G=e^{-2\psi}R'^2$.
Here $F(r,t)$ is the Misner-Sharp mass of the system describing
the amount of matter enclosed
within the shell labeled by $r$ at the time $t$.
\begin{equation}\label{Misner}
    F=R(1-G+e^{-2\nu}\dot{R}^2) \; .
\end{equation}
Note that equation \eqref{nu} together with the definition
of the scalar field implies that $\phi$ must obey the
Klein-Gordon equation. The system of Einstein equations
together with the equation of state and the Misner-Sharp
mass is a set of six equations in the six unknowns $\rho$, $p$,
$\psi$, $\nu$, $R$ and $F$, and is thus a fully closed
system for this case.

There is a scaling degree of freedom that we
can use to set $R(r,t_i)=r$. We then write $R(r,t)=rv(r,t)$
and rewrite all the equations in terms of the variable
$v$. Thus $v(r,t)$ is such that $v=1$
corresponds to the initial time and $v=0$ to the
time of occurrence of the singularity,
that is, the shell labeled by $r$ becomes
singular at the time $t_s(r)$ given by $v(r,t_s(r))=0$
(\cite{JG}, \cite{JD}).
The condition for continual collapse is
then given by $\dot{v}<0$, while the situation where
$\dot{v}$ changes sign at a finite time corresponds
to a halting and dispersal of the scalar field.
We can rewrite all the relevant functions in terms
of the new coordinates $\{r,v\}$, therefore for any
function $X(r,t)$ we have $X'=X_{,r}+X_{,v}v'$ where $v'(r,t)$
itself is treated as a function of $r$ and $v$.

Avoidance of shell-crossing singularities, which are
generally believed to be a breakdown of the coordinate
system rather than true singularities, is obtained if
we impose $R'>0$, which corresponds to $v+rv'>0$. Since
$v$ is always positive, there is always a small neighborhood
of the center $r=0$ for which no shell-crossing occurs,
and our analysis holds good for such a collapsing cloud.

In order to describe physically valid scenarios the energy
conditions must be fulfilled throughout collapse. In
this case these are always satisfied once $\rho$ is
taken to be positive, which,
provided $R'>0$, corresponds to $F'>0$.
Furthermore regularity must be imposed on the initial
data in order for
$\rho$ and the metric functions to be well behaved.
We shall require that the Misner-Sharp mass behaves
like $F=r^3M(r,v)$ with $M'(0,v)=0$ in order for the matter
density to be regular and without cusps at the center
at all epochs of collapse earlier than the
singularity.

An important point here is, to solve the system of
Einstein equations we
write all equations in terms of the mass function
$M(r,v)$ and its derivatives. From equation \eqref{nu}
we get $R^2e^{\psi-\nu}\dot{\phi}=f(r)$ where $f(r)$ is
an arbitrary function related to the kinetic energy of
the collapsing cloud.
From the definition of $\rho$ given by the energy-momentum
tensor we can write $e^{2\nu}=-{v^2\dot{\phi}^2}/{(2M_{,v})}$,
while using the equation of state
and the above definition for $f$ allows us to write $G$ as
$ G(r,v)=-v^2 {(vM_{,v}-3M-rM_{,r})^2}/{2f^2M_{,v}}$.
Now using equation \eqref{rho} and \eqref{p} we can write
the differential equation for $v'$ as
\begin{equation}
    v'=W(r,v)= -\frac{vM_{,v}+3M+rM_{,r}}{2M_{,v}} \; ,
\label{vdash}
\end{equation}
while from the Misner equation \eqref{Misner} we obtain
\begin{equation}
    \dot{v}=V(r,v)=-e^\nu\sqrt{\frac{M}{v}+\frac{G-1}{r^2}} \; ,
\label{vdot}
\end{equation}
with negative sign taken in order to describe collapse.
From the above we see that for collapse to occur we must
require a `reality condition', namely $\frac{M}{v}+\frac{G-1}{r^2}>0$.
If this condition is not satisfied throughout the evolution
then the system will reach $\dot{v}=0$ in a finite time and
collapse will then reverse to dispersal.
Finally the whole system reduces to a second order partial
differential equation in $M$ and its derivatives,
which can be written as
\begin{equation}\label{master}
    V_{,v}W-VW_{,v}=V_{,r} \; .
\end{equation}
This equation gives the integrability condition of
(\ref{vdash}) and (\ref{vdot})
that must be satisfied by the mass function $M(r,v)$ in
order for it to be a solution to the collapsing system.
Once a function $M(r,v)$ solving equation \eqref{master}
is found the
system of Einstein equations is fully solved.
It is clear that the above equation has a non-empty class
of solutions to which the self-similar collapse models
as well as the FLRW class belong.
By integrating the equation \eqref{vdot} we obtain
the function $t(r,v)$ that describes the time at which the shell
labeled by $r$ reaches the `epoch' $v$. Then we can
retrieve the singularity curve $t_s(r)$, namely the curve
describing the time at which each shell becomes singular,
by setting $t_s(r)=t(r,0)$.

For all sufficiently regular solutions $M(r,v)$
the singularity curve is also regular and can be written
near the center as,
\begin{equation}\label{ts}
    t_s(r)=t_0+\chi_1r+\chi_2r^2+o(r^3) \; ,
\end{equation}
where $t_0=t(0,0)$  is the time at which the central shell
becomes singular and
$\chi_i=\frac{1}{i!}\frac{d^it}{dr^i}|_{r=0}$.
Here $\chi_1$ vanishes due to regularity requirements.
The tangent to the singularity curve close to the center
is then determined by the coefficient $\chi_2$ and it is
then possible to show that it is this coefficient that
determines the local visibility of the central singularity
or otherwise
(\cite{JG}, \cite{Joshi}).
Specifically, if the singularity curve is an
increasing function of the coordinate $r$ at the center,
that is $\chi_2>0$, we then have an outgoing
null geodesic family coming out from the central singularity.
In that case the singularity is locally naked.
On the other hand, the singularity is covered if
$\chi_2\le 0$.

The FLRW collapse is obtained when we take
$M= {m_0}/{v^3}$, for which $v=v(t)$
and $G$ becomes $\frac{6m_0}{f^2}$ (from which we can see
that imposing the regularity condition $f^2(0)=6m_0$ ensures
$G=1$ at the center).
The singularity curve is constant in this case, all
shells fall into the singularity at the same time and the
event horizon forms before the time of formation of
the singularity, thus giving rise to a black hole
final state for collapse.
Going to more general case, solving the complete
second order PDE
\eqref{master} is in general unattainable.
Nevertheless our purpose here is to extract the crucial
information regarding the local visibility or otherwise
of the central
singularity, and this can be achieved by considering
a close neighborhood of the center. In order to do so
we expand Einstein equations and all relevant
functions in powers of $r$ close to the origin.

We therefore studied here a class of spacetimes obtained
by this procedure and the two-dimensional metric can be
given as,
\begin{equation}
    ds^2=-\left(1+\frac{m_{,v}}{v^2}r^2\right)dt^2
+v^2\left(1+\frac{m_{,v}v^4-5mv^3}{3m_0}r^2\right)dr^2. \;
\end{equation}
This line element is a solution of the Einstein
equations to second
order in $r$ which is valid in a small neighborhood of
$r=0$. The metric corresponds to a mass function
$M(r,v)=\frac{m_0}{v^3}+m(v)r^2$, a velocity profile
$f^2=6m_0$, and $\dot{\phi}(t(v,r))^2=\frac{6m_0}{v^6}$.
Here $m(v)$ is a solution to the corresponding
ordinary differential equation coming from the expansion of
equation \eqref{master}
to second order in $r$, which now becomes
\begin{eqnarray}\nonumber
    0&=&\left(\frac{5}{3}mv^9+m_0^2v^2\right)m_{,vv}
+5(mv^8+2m_0^2v)m_{,v}+ \\ \label{ODE}
    &&+\frac{50}{3}m^2v^7+24m_0^2m-m_{,v}^2v^9 \; ,
\end{eqnarray}
note that the zeroth order gives the FLRW differential
equation.

Once \eqref{ODE} is solved we obtain $m(v)$ from which
we can evaluate the singularity curve
as in equation \eqref{ts} to determine the coefficient
$\chi_2$, that eventually decides the local
visibility or otherwise of the singularity.
For the above class of solutions
$\chi_2$ is given by
\begin{equation}
    \chi_2=-\frac{1}{6}\int^1_0\frac{\frac{3m}{v}+m_{,v}
+\frac{5mm_{,v}v^7}{3m_0^2}+\frac{(5mv^3+m_{,v}v^4)^2}{12m_0^2}}
    {\left(\frac{m_0}{v^4}+\frac{5mv^3}{3m_0}\right)^{\frac{3}{2}}}dv \; .
\end{equation}

We note that the `reality condition' in this case takes
the form $\frac{m_0}{v^4}+\frac{5mv^3}{3m_0}>0$
and appears at the denominator in $\chi_2$. Therefore those
mass profiles $m$ for which it is
violated will have complex values for $\chi_2$.
Hence we can easily see that
all possible local behaviours of the central singularity
are entirely determined by the value of $\chi_2$. The central
singularity is locally naked if $\chi_2$ is positive,
is covered if it is negative, and there is a dispersal
of the collapsing central shell if $\chi_2$ is complex.

Given a certain initial data in the form
$\{m(1),m_{,v}(1)\}$, a solution of equation \eqref{ODE}
exists and it is possible
to solve it numerically and obtain the value for $\chi_2$.
If we choose initial data sufficiently close to FLRW,
meaning with $|m(1)|<m_0$, and require $m_{,v}(1)$
to be small to satisfy the `reality condition', we
can then evaluate $\chi_2$ as above and its sign will
then decide the visibility of the central singularity
for the specific massless scalar field collapse
given by that $m(v)$.

\begin{figure}[hhhh]
\includegraphics[scale=0.75]{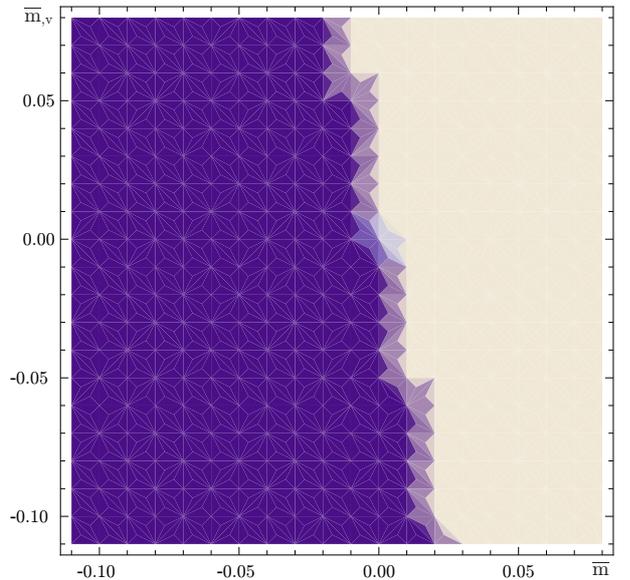}
\caption{The final outcome of collapse of a massless scalar
field depending on the initial data
$\{\bar{\text{m}}=m(1), \bar{\text{m}}_{,v}=m_{,v}(1)\}$ in a
neighborhood of the initial data for FLRW collapse.
For every point in the phase space of initial data, the sign of
$\chi_2$ was evaluated numerically.
The thick curve represents the boundary at which $\chi_2=0$
that separates the initial data leading
to a black hole outcome (on the left) from the initial
data leading to a naked singularity (on the right).
Note that $m(1)=m_{,v}(1)=0$  corresponds to initial
data for the FLRW solution.}
\label{fig}
\end{figure}

What we find then is, for initial data close to
FLRW, both the collapse outcomes namely the black hole
and naked singularity are equally possible and the space
of initial data leading to either of them within this
specific class is evenly divided (see Fig. \ref{fig}).
The general behaviour found here for the initial data sets
close to FLRW resembles the results that were found
in similar scenarios for collapse of perfect fluids
and fluids sustained only by tangential pressure
\cite{MJ}.
The main virtue of the study of scalar field collapse comes
from the fact that these are
fundamental matter fields obeying the Klein-Gordon equation
and therefore are possibly valid at all late stages of
collapse, if quantum gravity effects are ignored.
Also, such a `stiff fluid' is a fully closed system where
there are no global free functions, as in the case of
perfect fluids or tangential pressures when no equations
of state are specified. So the
physical meaning of the matter model comes out in
a clear and straightforward manner.

The above result provides intriguing perspective
on the genericity aspect of occurrence of black holes and
naked singularity final states in gravitational collapse.
The definition of genericity in general relativity is
a delicate matter that requires a proper and deeper
understanding of both the measure and topology of the
parent space than is available presently. No precisely
well-defined criteria are available for the same in the
gravitation theory today. What we have showed here is
that there exist classes of solutions for which initial
data arbitrarily close to that of the FLRW models
lead the collapse to a naked singularity. Furthermore,
the set of initial data leading to a naked singularity
within this particular space of solutions does not reduce
to a set of zero measure, and the same is true for the
black hole outcomes. In this sense, within this class,
the naked singularity formation can be considered
somehow generic.

Finally we would like to emphasize that what
we deduced here is the occurrence of a
locally naked singularity that disproves the
stronger version of the censorship conjecture. It is
possible in principle that the singularity is only
locally naked and trajectories do come out, but then
they eventually all fall back into the horizon,
without being globally visible, and the weaker version
of the censorship conjecture may still be obeyed.
This needs to be examined separately.

Acknowledgement: The authors would like to thank Mandar Patil
for the help with Fig 1. RG would like to thank
Claude Leon Foundation Postdoctoral Fellowship, South Africa.

\end{document}